\documentclass[twocolumn]{aastex631}

\newcommand{\kms}{{\rm km~s^{-1}}}
\newcommand{\oii}{[\textrm{O}~\textsc{ii}]}
\newcommand{\oiii}{[\textrm{O}~\textsc{iii}]}

\newcommand{\cii}{[\textrm{C}~\textsc{ii}]}
\newcommand{\hi}{\textrm{H}~\textsc{i}}
\newcommand{\hii}{\textrm{H}~\textsc{ii}}

\newcommand{\simgt}{\,\rlap{\lower 3.5 pt \hbox{$\mathchar \sim$}} \raise
1pt \hbox {$>$}\,}
\newcommand{\simlt}{\,\rlap{\lower 3.5 pt \hbox{$\mathchar \sim$}} \raise
1pt \hbox {$<$}\,}

\newcommand{\ly}{${\rm Ly\alpha}$}

\newcommand{\hb}{${\rm H\beta}$}

\newcommand{\Z}{{\rm [Z/H]}}

\newcommand{\grizli}{{\tt Grizli}}
\newcommand{\gsf}{{\tt gsf}}
\newcommand{\targ}{A2744-z7p9OD}
\newcommand{\nspec}{seven}

\submitjournal{ApJ Letters}

\shorttitle{Spectroscopic confirmation of a protocluster at $z=7.88$}
\shortauthors{Morishita et al.}
\graphicspath{{./}{figures/}}
\begin{document}

\title{ 
Early results from GLASS-JWST. XIV: A spectroscopically confirmed protocluster 650 million years after the Big Bang
}

\correspondingauthor{Takahiro Morishita}
\email{takahiro@ipac.caltech.edu}

\author[0000-0002-8512-1404]{Takahiro Morishita}
\affiliation{IPAC, California Institute of Technology, MC 314-6, 1200 E. California Boulevard, Pasadena, CA 91125, USA}

\author[0000-0002-4140-1367]{Guido Roberts-Borsani}
\affiliation{Department of Physics and Astronomy, University of California, Los Angeles, 430 Portola Plaza, Los Angeles, CA 90095, USA}

\author[0000-0002-8460-0390]{Tommaso Treu}
\affiliation{Department of Physics and Astronomy, University of California, Los Angeles, 430 Portola Plaza, Los Angeles, CA 90095, USA}

\author[0000-0003-2680-005X]{Gabriel Brammer}
\affiliation{Cosmic Dawn Center (DAWN)}
\affiliation{Niels Bohr Institute, University of Copenhagen, Jagtvej 128, DK-2200 Copenhagen N, Denmark}

\author[0000-0002-8460-0390]{Charlotte A. Mason}
\affiliation{Cosmic Dawn Center (DAWN)}
\affiliation{Niels Bohr Institute, University of Copenhagen, Jagtvej 128, DK-2200 Copenhagen N, Denmark}

\author[0000-0001-9391-305X]{Michele Trenti}
\affiliation{School of Physics, University of Melbourne, Parkville 3010, VIC, Australia}
\affiliation{ARC Centre of Excellence for All Sky Astrophysics in 3 Dimensions (ASTRO 3D), Australia}

\author[0000-0003-0980-1499]{Benedetta Vulcani}
\affiliation{INAF Osservatorio Astronomico di Padova, vicolo dell'Osservatorio 5, 35122 Padova, Italy}

\author[0000-0002-9373-3865]{Xin Wang}
\affil{School of Astronomy and Space Science, University of Chinese Academy of Sciences (UCAS), Beijing 100049, China}
\affil{National Astronomical Observatories, Chinese Academy of Sciences, Beijing 100101, China}
\affil{Institute for Frontiers in Astronomy and Astrophysics, Beijing Normal University,  Beijing 102206, China}

\author[0000-0003-3108-9039]{Ana~Acebron}
\affiliation{Dipartimento di Fisica, Università degli Studi di Milano, Via Celoria 16, I-20133 Milano, Italy}
\affiliation{INAF - IASF Milano, via A. Corti 12, I-20133 Milano, Italy}

\author[0000-0002-3196-5126]{Yannick Bah\'{e}}
\affiliation{Leiden Observatory, Leiden University, P.O. Box 9513, 2300 RA Leiden,
The Netherlands}

\author[0000-0003-1383-9414]{Pietro~Bergamini}
\affiliation{Dipartimento di Fisica, Università degli Studi di Milano, Via Celoria 16, I-20133 Milano, Italy}
\affiliation{INAF - OAS, Osservatorio di Astrofisica e Scienza dello Spazio di Bologna, via Gobetti 93/3, I-40129 Bologna, Italy}

\author[0000-0003-4109-304X]{Kristan~Boyett}
\affiliation{School of Physics, University of Melbourne, Parkville 3010, VIC, Australia}
\affiliation{ARC Centre of Excellence for All Sky Astrophysics in 3 Dimensions (ASTRO 3D), Australia}

\author[0000-0001-5984-0395]{Marusa Bradac}
\affiliation{University of Ljubljana, Department of Mathematics and Physics, Jadranska ulica 19, SI-1000 Ljubljana, Slovenia}
\affiliation{Department of Physics and Astronomy, University of California Davis, 1 Shields Avenue, Davis, CA 95616, USA}

\author[0000-0003-2536-1614]{Antonello Calabr\`o}
\affiliation{INAF Osservatorio Astronomico di Roma, Via Frascati 33, 00078 Monteporzio Catone, Rome, Italy}

\author[0000-0001-9875-8263]{Marco Castellano}
\affiliation{INAF Osservatorio Astronomico di Roma, Via Frascati 33, 00078 Monteporzio Catone, Rome, Italy}

\author[0000-0003-1060-0723]{Wenlei Chen}
\affiliation{School of Physics and Astronomy, University of Minnesota, 116 Church Street SE, Minneapolis, MN 55455, USA}

\author[0000-0002-6220-9104]{Gabriella De Lucia}
\affiliation{INAF-Astronomical Observatory of Trieste, via G.~B. Tiepolo 11, I-34143, Trieste, Italy}

\author[0000-0003-3460-0103]{Alexei V. Filippenko}
\affiliation{Department of Astronomy, University of California, Berkeley, CA 94720-3411}

\author[0000-0003-3820-2823]{Adriano~Fontana}
\affiliation{INAF Osservatorio Astronomico di Roma, Via Frascati 33, 00078 Monteporzio Catone, Rome, Italy}

\author[0000-0002-3254-9044]{Karl Glazebrook}
\affiliation{Centre for Astrophysics and Supercomputing, Swinburne University of Technology, PO Box 218, Hawthorn, VIC 3122, Australia}

\author[0000-0002-5926-7143]{Claudio~Grillo}
\affiliation{Dipartimento di Fisica, Università degli Studi di Milano, Via Celoria 16, I-20133 Milano, Italy}
\affiliation{INAF - IASF Milano, via A. Corti 12, I-20133 Milano, Italy}

\author[0000-0002-6586-4446]{Alaina Henry}
\affiliation{Center for Astrophysical Sciences, Department of Physics \& Astronomy, Johns Hopkins University, Baltimore, MD 21218, USA}
\affiliation{Space Telescope Science Institute, 3700 San Martin Drive, Baltimore, MD 21218, USA}

\author[0000-0001-5860-3419]{Tucker Jones}
\affiliation{Department of Physics and Astronomy, University of California Davis, 1 Shields Avenue, Davis, CA 95616, USA}

\author[0000-0003-3142-997X]{Patrick L. Kelly}
\affiliation{Minnesota Institute for Astrophysics, University of Minnesota, 116 Church Street SE, Minneapolis, MN 55455, USA}

\author[0000-0002-6610-2048]{Anton M. Koekemoer}
\affiliation{Space Telescope Science Institute, 3700 San Martin Dr., 
Baltimore, MD 21218, USA}

\author[0000-0003-4570-3159]{Nicha Leethochawalit}
\affiliation{National Astronomical Research Institute of Thailand (NARIT), Mae Rim, Chiang Mai, 50180, Thailand}

\author[0000-0002-4965-6524]{Ting-Yi Lu}
\affiliation{Cosmic Dawn Center (DAWN)}
\affiliation{Niels Bohr Institute, University of Copenhagen, Jagtvej 128, DK-2200 Copenhagen N, Denmark}

\author[0000-0001-9002-3502]{Danilo Marchesini}
\affiliation{Physics and Astronomy Department, Tufts University, 574 Boston Avenue, Medford, MA 02155, USA}

\author[0000-0002-9572-7813]{Sara Mascia}
\affiliation{INAF Osservatorio Astronomico di Roma, Via Frascati 33, 00078 Monteporzio Catone, Rome, Italy}

\author[0000-0001-9261-7849]{Amata Mercurio}
\affiliation{INAF -- Osservatorio Astronomico di Capodimonte, Via Moiariello 16, I-80131 Napoli, Italy}

\author[0000-0001-6870-8900]{Emiliano Merlin}
\affiliation{INAF Osservatorio Astronomico di Roma, Via Frascati 33, 00078 Monteporzio Catone, Rome, Italy}

\author[0000-0002-8632-6049]{Benjamin~Metha}
\affiliation{School of Physics, University of Melbourne, Parkville 3010, VIC, Australia}
\affiliation{ARC Centre of Excellence for All Sky Astrophysics in 3 Dimensions (ASTRO 3D), Australia}
\affiliation{Department of Physics and Astronomy, University of California, Los Angeles, 430 Portola Plaza, Los Angeles, CA 90095, USA}

\author[0000-0003-2804-0648 ]{Themiya Nanayakkara}
\affiliation{Centre for Astrophysics and Supercomputing, Swinburne University of Technology, PO Box 218, Hawthorn, VIC 3122, Australia}

\author[0000-0001-6342-9662]{Mario Nonino}
\affiliation{INAF-Trieste Astronomical Observatory, Via Bazzoni 2, I-34124, Trieste, Italy}

\author[0000-0002-7409-8114]{Diego Paris}
\affiliation{INAF Osservatorio Astronomico di Roma, Via Frascati 33, 00078 Monteporzio Catone, Rome, Italy}

\author[0000-0001-8940-6768 ]{Laura Pentericci}
\affiliation{INAF Osservatorio Astronomico di Roma, Via Frascati 33, 00078 Monteporzio Catone, Rome, Italy}

\author[0000-0002-6813-0632]{Piero~Rosati}
\affiliation{Dipartimento di Fisica e Scienze della Terra, Università degli Studi di Ferrara, Via Saragat 1, I-44122 Ferrara, Italy}
\affiliation{INAF - OAS, Osservatorio di Astrofisica e Scienza dello Spazio di Bologna, via Gobetti 93/3, I-40129 Bologna, Italy}

\author[0000-0002-9334-8705]{Paola Santini}
\affiliation{INAF Osservatorio Astronomico di Roma, Via Frascati 33, 00078 Monteporzio Catone, Rome, Italy}

\author[0000-0002-6338-7295]{Victoria Strait}
\affiliation{Cosmic Dawn Center (DAWN)}
\affiliation{Niels Bohr Institute, University of Copenhagen, Jagtvej 128, DK-2200 Copenhagen N, Denmark}

\author[0000-0002-5057-135X]{Eros~Vanzella}
\affiliation{INAF -- OAS, Osservatorio di Astrofisica e Scienza dello Spazio di Bologna, via Gobetti 93/3, I-40129 Bologna, Italy}

\author[0000-0001-8156-6281]{Rogier A. Windhorst} %%% Rogier.Windhorst@gmail.com
\affiliation{School of Earth and Space Exploration, Arizona State University, Tempe, AZ 85287-1404, USA}

\author[0000-0003-3864-068X]{Lizhi Xie}
\affiliation{Tianjin Normal University, Binshuixidao 393, 300387, Tianjin, China}
\affiliation{INAF-Astronomical Observatory of Trieste, via G.~B. Tiepolo 11, I-34143, Trieste, Italy}



\begin{abstract}
We present the spectroscopic confirmation of a protocluster at $z=7.88$ behind the galaxy cluster Abell~2744 (hereafter \targ). Using JWST NIRSpec, we find \nspec\ galaxies within a projected radius of 60\,kpc. Although the galaxies reside in {an overdensity around $\simgt20\times$ greater than a random volume}, they do not show strong \ly\ emission.  We place 2-$\sigma$ upper limits on the rest-frame equivalent width $<16$--$28$\,\AA.
Based on the tight upper limits to the \ly\ emission, we constrain the {volume-averaged neutral fraction of hydrogen in the intergalactic medium} to be $x_{\rm HI} > 0.45$ (68\,\% CI). Using an empirical $M_{\rm UV}$-$M_{\rm halo}$ relation for individual galaxies, we estimate that the total halo mass of the system is $\gtrsim 4\times10^{11}\,M_\odot$. Likewise, the line of sight velocity dispersion is estimated to be $1100\pm200\,\kms$. {Using an empirical relation, we estimate the present-day halo mass of \targ\ to be $\sim2\times10^{15}\,M_\odot$, comparable to the Coma cluster.}
\targ\ is the highest redshift spectroscopically confirmed protocluster to date, demonstrating the power of JWST to investigate the connection between dark-matter halo assembly and galaxy formation at very early times with medium-deep observations at $<20$\,hr total exposure time. 
Follow-up spectroscopy of the remaining photometric candidates of the overdensity will further refine the features of this system and help characterize the role of such overdensities in cosmic reionization. 
\end{abstract}

%Classical Novae (251) --- Ultraviolet astronomy(1736) --- History of astronomy(1868) --- Interdisciplinary astronomy(804)}


\section{Introduction} \label{sec:intro}
Hierarchical structure formation is one of the fundamental features of our standard cosmological model. The first overdensities to collapse and form stars and galaxies play a particularly important role in the evolution of the universe and cosmic reionization \citep{tegmark1997}. Identifying and studying the sources associated with these first overdensities thus provides critical insights into the evolution of galaxies, the intergalactic medium, and the underlying dark matter scaffolding \citep[e.g.,][]{mo96}. 

The clustering of sources around a luminous galaxy or quasar includes an expected excess of fainter companions, under a broad assumption that galaxy luminosity is correlated with the mass of the dark-matter host halo. Such early overdensities are thought to be the seeds of today's galaxy clusters, and sites where galaxy formation and the evolution of the surrounding gas is progressing more rapidly compared to the mean of the universe. As such, the identification of galaxy over-densities at high redshift ($z>6$) has been of particular interest in the literature \citep[e.g.,][]{trenti12,castellano16,castellano18,castellano22b,harikane19,tilvi20,hu21,endsley22,larson22}. Furthermore, galaxy overdensities serve as ideal laboratories for studying the ionization of neutral hydrogen around galaxy systems; the presence of a large ionizing bubble may boost the fraction of escaping \ly\ photons, which otherwise are scattered and absorbed by surrounding neutral hydrogen \citep[][see also \citealt{trapp22}]{miralda-escude98,dijkstra14,mason20}. 

An excess of photometric $z\sim8$ sources behind the massive galaxy cluster Abell~2744 was discovered in deep HST images taken as part of the Hubble Frontier Fields program \citep{lotz17} and has been extensively investigated since \citep[][]{zheng14,atek15,ishigaki16}. Approximately a dozen photometrically-selected sources are distributed within a small region ($\sim20''$ across), making it an extreme over-density, with
$\delta\sim130_{-51}^{+66}$ \citep{ishigaki16}, where $\delta=(n-\bar{n})/\bar{n}$ represents the excess of surface number density from the field average. 

Spectroscopic follow-up of a number of those sources with VLT/X-Shooter, ALMA, and JWST/NIRISS has secured spectroscopic redshifts for three sources at $z>7$ \citep{laporte17,laporte19,carniani20,roberts-borsani22}. Of particular interest is the Lyman-break galaxy, YD4, a photometrically-selected candidate member of the overdensity which revealed Lyman-$\alpha$, \oiii\ 88 $\mu$m at $z=8.38$ as well as the presence of dust in its proximity \citep[][but see Sec.~\ref{ssec:previous}]{laporte17}.

%%%%%%%%%%%%%%%%%%%%%
\begin{figure*}
\centering
	\includegraphics[width=0.8\textwidth]{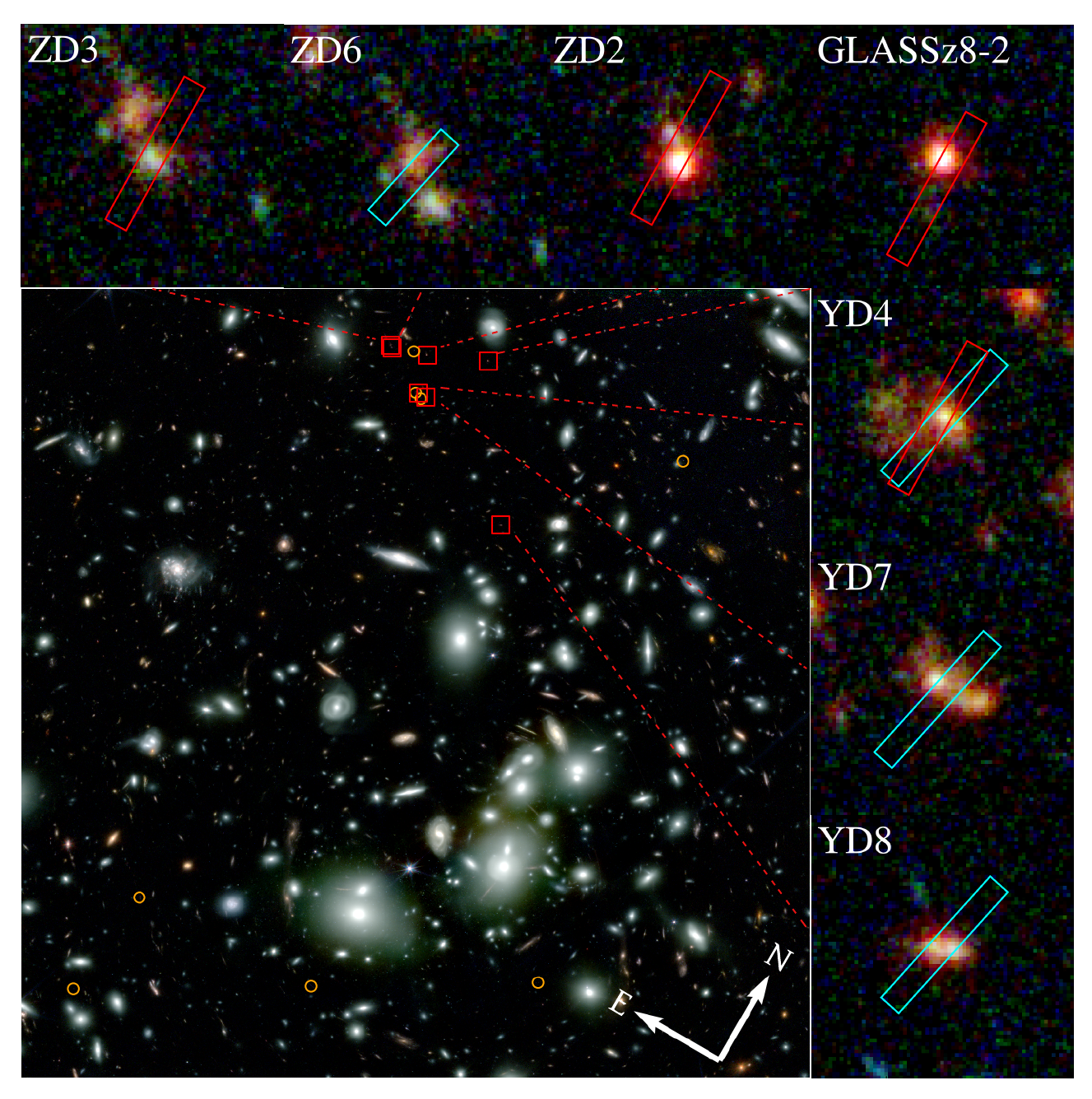}
	\caption{
	NIRCam RGB composite image of the Abell~2744 field (blue:F115W, green:F200W, red:F444W). Confirmed galaxies are marked by red squares and shown individually in the zoomed-in panels ($2.\!''2\times2.\!''2$). The position of the MSA slit for each object is shown a rectangle (colored in cyan for the DDT program and red for the GLASS-ERS). The remaining photometric $z\sim7.9$ candidates that were originally identified in \citet{zheng14} but not covered by our NIRSpec observations, are marked in orange circles.
    }
\label{fig:mosaic}
\end{figure*}
%%%%%%%%%%%%%%%%%%%%%

Here we report the spectroscopic follow-up and confirmation of the overdensity at $z=7.88$ (hereafter \targ; {Figure~\ref{fig:mosaic}}), through the detection of the \oiii$\lambda$5007 line in \nspec\ member galaxies with JWST/NIRSpec. This result is consistent with the hypothesis that the galaxy confirmed by previous work at $z=8.38$ is in the background of the protocluster identified here, highlighting the importance of spectroscopic confirmation to establish membership and overdensity, accounting for chance alignment of galaxies sharing similar photometric redshifts.

The cluster field Abell~2744 is the primary target of the GLASS-JWST Early Release Science program \citep[JWST-GO-1324;][P.I. Treu]{treu22,roberts-borsani22}, and also part of the JWST Director Discretionary Time program \cite[JWST-GO-2756; P.I. Chen;][]{roberts-borsani22b} to follow up the discovery of a magnified supernova at $z=3.47$ \citep{chen22}. The wavelength coverage $0.6$--$5.3\,\mu$m afforded by the NIRSpec observations not only allows for redshift confirmations  of the candidate members via a large suite of emission lines, but also provides insight into the visibility of \ly\ from galaxies in an overdense region. The unique data set is complemented by deep JWST/NIRCam and ancillary HST images, allowing us to characterize the physical properties of the confirmed members and infer the early evolution of galaxies in such an extreme environment. Furthermore, a first estimate of the velocity dispersion of the protocluster can be derived based on the high-precision redshift measurements for individual members. 

The paper is structured as follows: we present the data set in Sec.~\ref{sec:data}, followed by our spectroscopic and photometric analyses of the cluster members in Sec.~\ref{sec:ana}. We characterize the system and infer the neutral fraction around the system in Sec.~\ref{sec:disc}, and summarize our key conclusions in Sec.~\ref{sec:sum}. Where relevant, we adopt the AB magnitude system \citep{oke83,fukugita96}, cosmological parameters of $\Omega_m=0.3$, $\Omega_\Lambda=0.7$, $H_0=70\,\kms\, {\rm Mpc}^{-1}$, and the \citet{chabrier03} initial mass function. Distances are in proper units unless otherwise stated.

%%%%%%%%%%%%%%%%%%%%%
\begin{deluxetable*}{ccccclccc}
\tablecaption{Spectroscopically confirmed protocluster member galaxies behind Abell~2744}
\tabletypesize{\footnotesize}
\tablecolumns{9}
\tablewidth{0pt} 
\tablehead{
\colhead{ID} & \colhead{R.A.} & \colhead{Decl.} & \colhead{m$_{\rm F150W}$} & \colhead{m$_{\rm F444W}$} & \colhead{Redshift} & \colhead{$f_{\rm H\beta}$} & \colhead{$f_{\rm{[OIII]}4959+5007}$} &  \colhead{EW$_0$(\ly)$^\dagger$}\\
\colhead{} & \colhead{degree} & \colhead{degree} & \colhead{mag} & \colhead{mag} & \colhead{} & \colhead{{$10^{-19}$\,erg/s/cm$^2$}}& \colhead{{$10^{-18}$\,erg/s/cm$^2$}} & \colhead{\AA}
}
\startdata
%\cutinhead{spec-z confirmed}
YD4 & 3.6038544 & -30.3822365 & 26.5 & 25.6 & $7.8758_{-0.0001}^{+0.0001}$ &$8.6_{-1.4}^{+1.6}$ & $20.2_{-0.7}^{+0.7}$ & $<15.9$\\
YD7 & 3.6033909 & -30.3822289 & 26.4 & 25.8 & $7.8772_{-0.0041}^{+0.0041}$ & $<76.3$ & $40.4_{-8.1}^{+8.8}$ & $<23.9$\\
ZD6 & 3.6065702 & -30.3808918 & 26.9 & 26.1 & $7.8843_{-0.0013}^{+0.0013}$ &$<13.6$ & $44.0_{-3.4}^{+3.3}$ & $<26.2$\\
YD8 & 3.5960841 & -30.3858051 & 26.7 & 26.0 & $7.8869_{-0.0004}^{+0.0004}$ &$109.0_{-13.2}^{+14.6}$ & $275.1_{-6.1}^{+6.2}$ & $<16.9$\\
ZD2 & 3.6045184 & -30.3804321 & 26.2 & 25.2 & $7.8800_{-0.0001}^{+0.0001}$ &$19.6_{-0.8}^{+0.8}$ & $70.1_{-0.5}^{+0.5}$ & --\\%$<17.5$
ZD3 & 3.6064637 & -30.3809624 & 26.7 & 26.6 & $7.8816_{-0.0001}^{+0.0001}$ &$4.5_{-0.7}^{+0.8}$ & $6.4_{-0.2}^{+0.2}$ & --\\
GLASSz8-2 & 3.6013467 & -30.3791904 & 26.5 & 25.4 & $7.8831_{-0.0001}^{+0.0001}$ &$8.2_{-0.6}^{+0.7}$ & $16.1_{-0.4}^{+0.4}$ & $<27.6$\\
\enddata
\tablecomments{
%{\bfr Ask Piero about negative magnif.}
$^\dagger$ 2\,$\sigma$ rest-frame equivalent width of \ly\ over $\Delta\sim100$\,\AA\ ($\sim2700\,\kms$).
\ly\ equivalent width measurements of ZD2 and ZD3 are not available as the \ly\ wavelength falls in the detector gap.
Measurements here are not corrected for magnification. 
{For those with \hb\ not detected, $2\,\sigma$ flux limit (assuming $100$\,\AA\ for the line width) are presented.}
}\label{tab:source}
\end{deluxetable*}

%%%%%%%%%%%%%%%%%%%%%
\section{Data}\label{sec:data}

\subsection{JWST/NIRSpec MSA observations}\label{sec:data-jwst}
We base our primary analysis on data acquired through NIRSpec MSA observations in two programs, the GLASS-JWST Early Release Science Program \citep[PID~1324, PI Treu;][]{treu22} and a JWST DDT program \citep[PID~2756, PI. W. Chen;][]{roberts-borsani22b}. The GLASS-JWST observations were executed on November 10, 2022 with three spectral resolution configurations with three high-resolution gratings, G140H/F100LP, G235H/F170LP, and G395H/F290LP, which also provide total wavelength coverage of 0.81--5.14\,$\mu$m, at $R\sim1000$\,--\,$3000$. The on-source exposure time was 4.9\,hours in each spectral configuration. The DDT NIRSpec observations were executed on October 23 2022, with the CLEAR filter+prism configuration, which provides continuous wavelength coverage of 0.6--5.3\,$\mu$m at $R\sim30$--300 spectral resolution. The on-source exposure time was 1.23\,hours. 

For the MSA target selection, we started with the same source catalog for both programs. Specifically for the $z\sim8$ protocluster sources, $z$/$Y$-dropout galaxies (hereafter ZDs and YDs, respectively) were included (4 in the DDT and 4 in GLASS-JWST), all within the vicinity of the overdensity \citep{zheng14,ishigaki16} including the spectroscopically-confirmed galaxies YD4, GLASSZ8-1 (ZD2) and GLASSZ8-2 from \citet{laporte17} and \citet{roberts-borsani22}, respectively. Considering the overlap between the two programs, a total of seven distinct protocluster targets were observed, but data was corrupted for one target due a non-nominal operation of a micro-shutter, leaving \nspec\ targets suitable for analysis. The choice of protocluster targets in each MSA was based on three primary factors, namely (i) the central pointing of the MSA, (ii) the position of the MSA  ensuring no spectral overlap in the detector, and (iii) preferential selection of brighter objects to maximize the probability of emission line or continuum detections.

The data were reduced using the official STScI JWST 
pipeline (ver.1.8.2)\footnote{\url{https://github.com/spacetelescope/jwst}} for Level 1 data products, and the \texttt{msaexp}\footnote{\url{https://github.com/gbrammer/msaexp}} code for Level 2 and 3 data products, the latter of which is built on the STScI pipeline routines but also includes custom routines for additional corrections. Briefly, we begin our data reduction on the uncalibrated files with the \texttt{Detector1Pipeline} routine and the latest set of reference files (jwst\_1014.pmap) to correct for detector-level artifacts and convert to count-rate images. We then utilize custom pre-processing routines from \texttt{msaexp} to correct for 1/$f$ noise, identify and remove ``snowballs'', and remove bias on an exposure-by-exposure basis, before running a number of STScI routines from the \texttt{Spec2Pipeline} to produce the final 2D cutout images. These include the \texttt{AssignWcs}, \texttt{Extract2dStep}, \texttt{FlatFieldStep}, \texttt{PathLossStep}, and \texttt{PhotomStep} routines to perform WCS registration, flat-fielding, pathloss corrections, and flux calibration. Background subtraction is performed locally using a three-shutter nod pattern before drizzling the resulting images onto a common grid. From there, we optimally extract the spectra via an inverse-variance weighted kernel, derived by summing the 2D spectrum along the dispersion axis and fitting the resulting signal along the spatial axis with a Gaussian profile by following the recipe of \citet[][]{horne86}. We visually inspect all kernels to ensure spurious events are not included (or limited) where possible. The kernel then extracts the 1D spectrum along the dispersion axis.

%%%%%%%%%%%%%%%%%%%%%
\subsection{Imaging data and photometry} 
\label{sec:hst}

Deep NIRCam images are available from DDT program (PID 2756; PI. W. Chen) and GO program UNCOVER \citep[][]{bezanson22}, including F115W, F150W, F200W, F277W, F356W, F410M, and F444W filters. The imaging data are reduced in the same way as presented by \citet{merlin22}, using the official STScI JWST pipeline, including the most recent version of the photometric zero points and reference files. {Images are PSF-matched to the F444W filter for the flux estimates below.}

To supplement our photometric wavelength coverage, we include ancillary \textit{Hubble} Space Telescope data taken by several programs \citep{postman12,treu15,lotz17,steinhardt20}. The HST data have been uniformly re-reduced using \grizli\ \citep{brammer22}. {The HST images are PSF-matched to the F160W filter instead of the NIRCam F444W. The choice was made because -- despite their similar PSF FWHMs -- there are significant differences in the PSF profile of the two different telescopes which make it challenging to obtain a satisfying convolution kernel. The remaining systematic offset in fluxes caused by this is corrected in the following process.}

A photometric catalog is constructed following \citet{morishita22}, using {\tt borgpipe} \citep{morishita21}. Briefly, fluxes are estimated in the PSF-matched  images with a $r=0.\!''32$ aperture by using SExtractor \citep{bertin96}. Flux offsets between NIRCam and HST filters are corrected with a rescaling based on the mean offset between NIRCam F150W and a pseudo F150W filter derived for the same sources using the HST F140W and F160W fluxes, whose broad band filters straddle the NIRCam F150W. {The correcting factor is found to be $1.25$, which is consistent with \citet{morishita22}.}

Lastly, fluxes are scaled to a total flux by applying $C = f_{\rm auto, F444W}/f_{\rm aper, F444W}$, where $f_{\rm auto, F444W}$ is {\rm FLUX\_AUTO} of SExtractor, measured for individual sources. 

%%%%%%%%%%%%%%%%%%%%%
\begin{figure*}
\centering
	\includegraphics[width=0.48\textwidth]{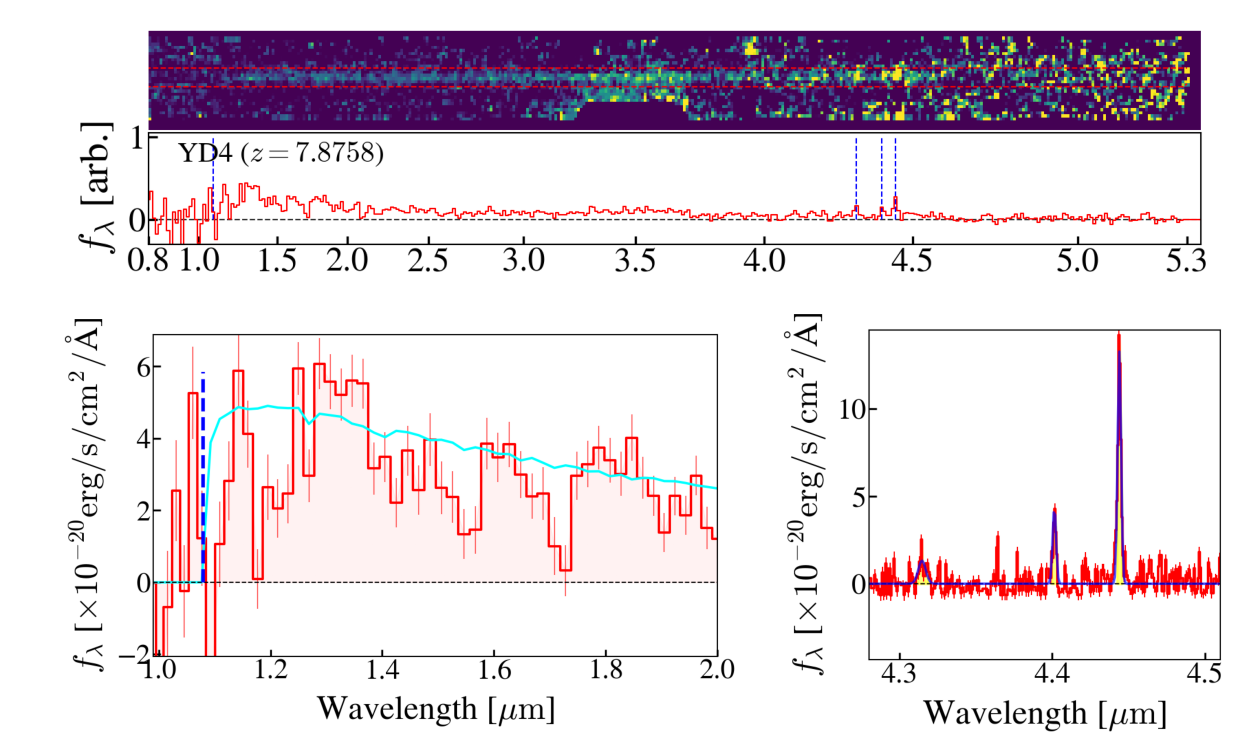}
	\includegraphics[width=0.48\textwidth]{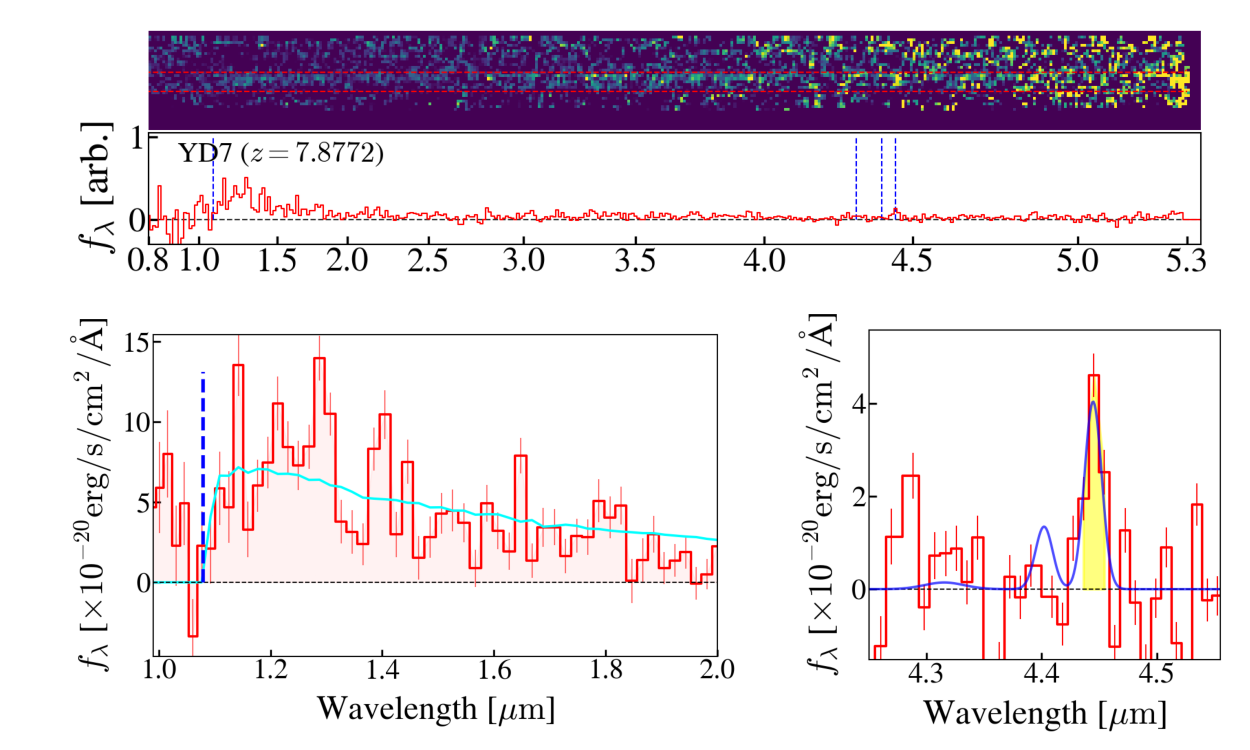}
	\includegraphics[width=0.48\textwidth]{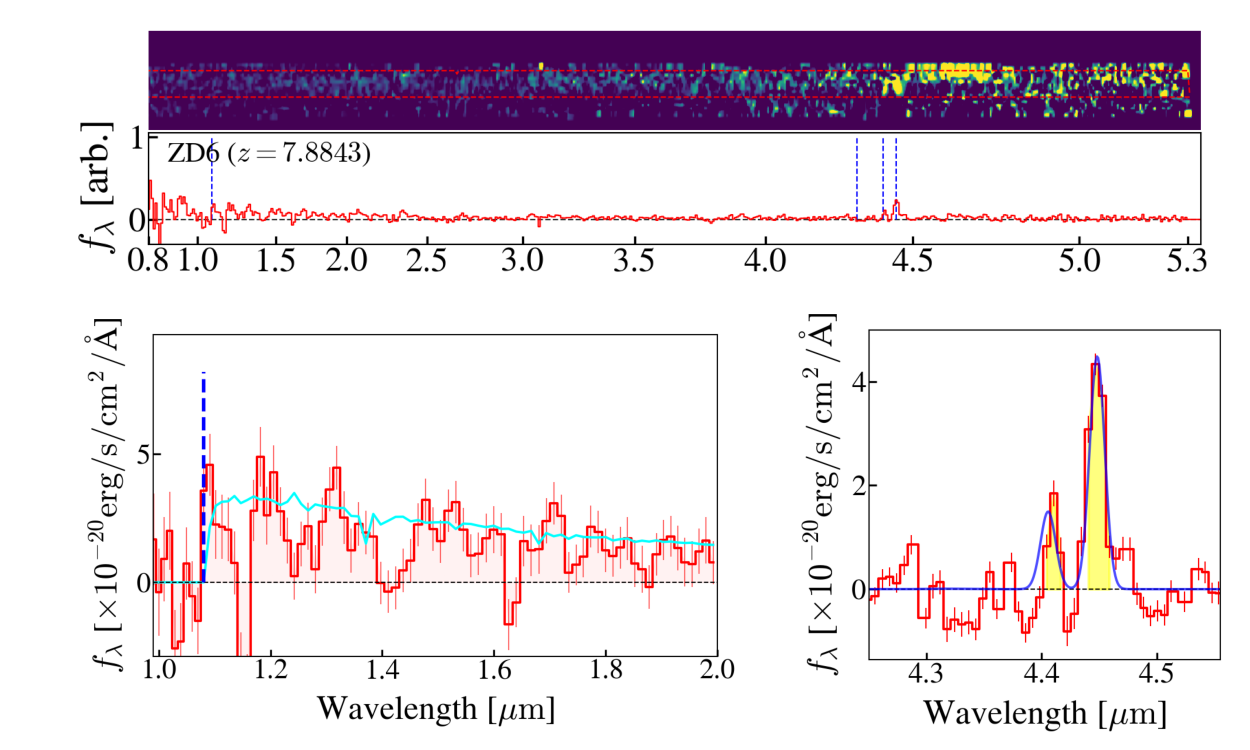}
	\includegraphics[width=0.48\textwidth]{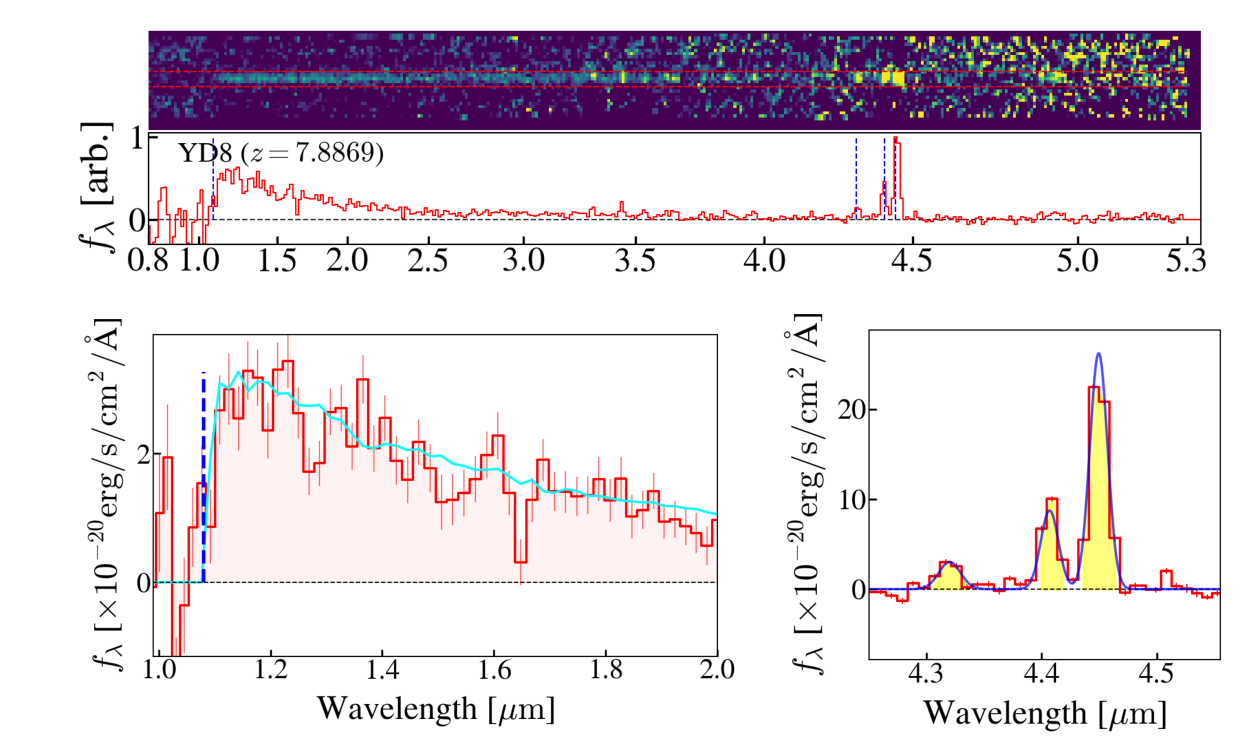}
	\includegraphics[width=0.48\textwidth]{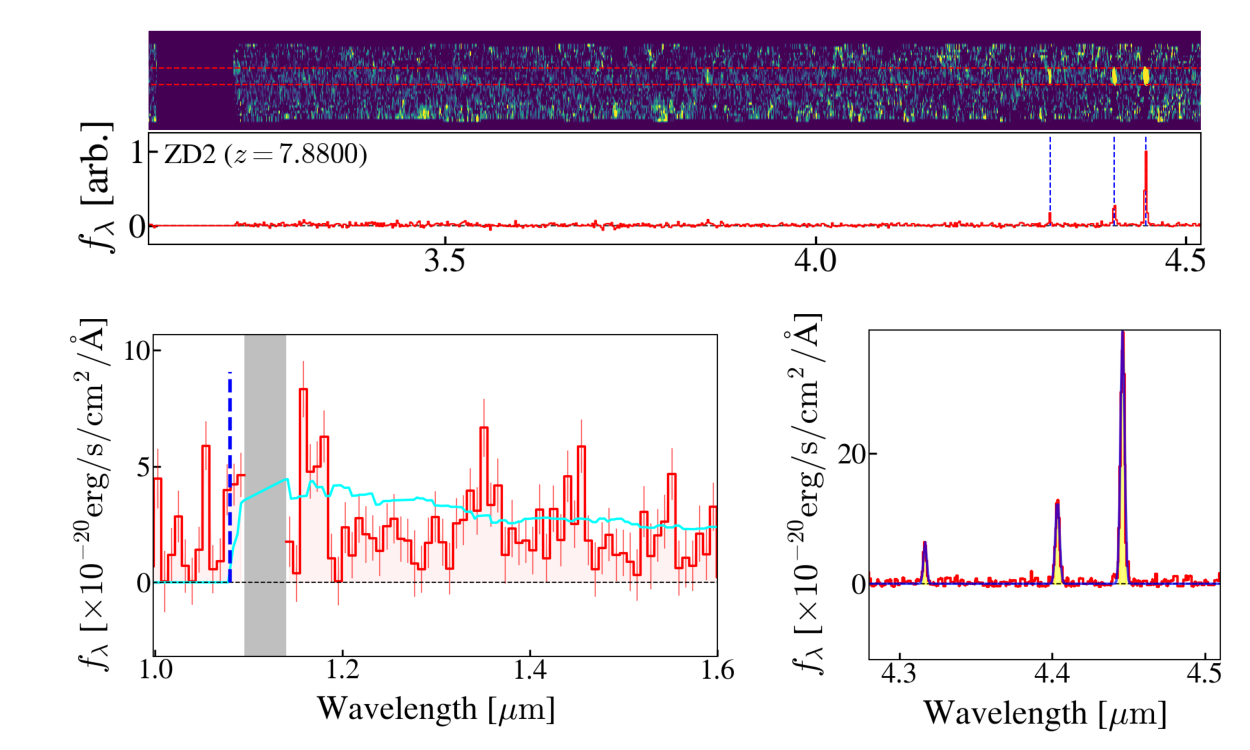}
	\includegraphics[width=0.48\textwidth]{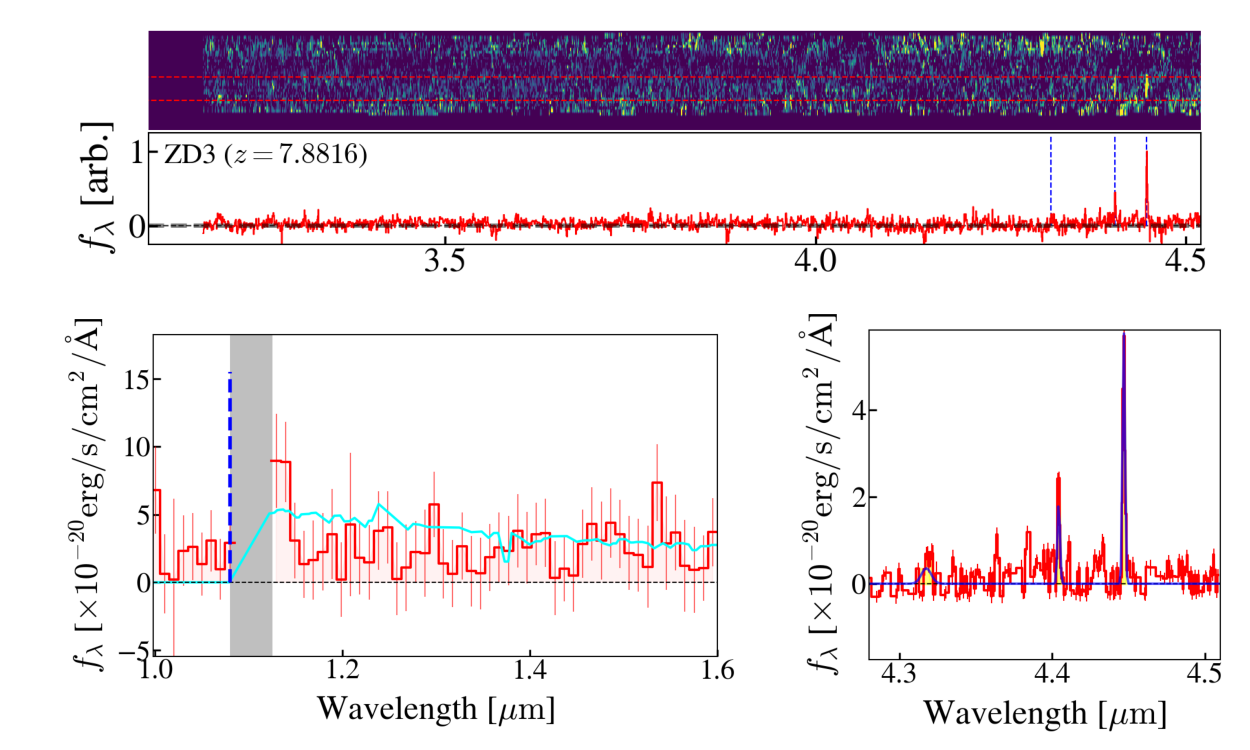}
	\includegraphics[width=0.48\textwidth]{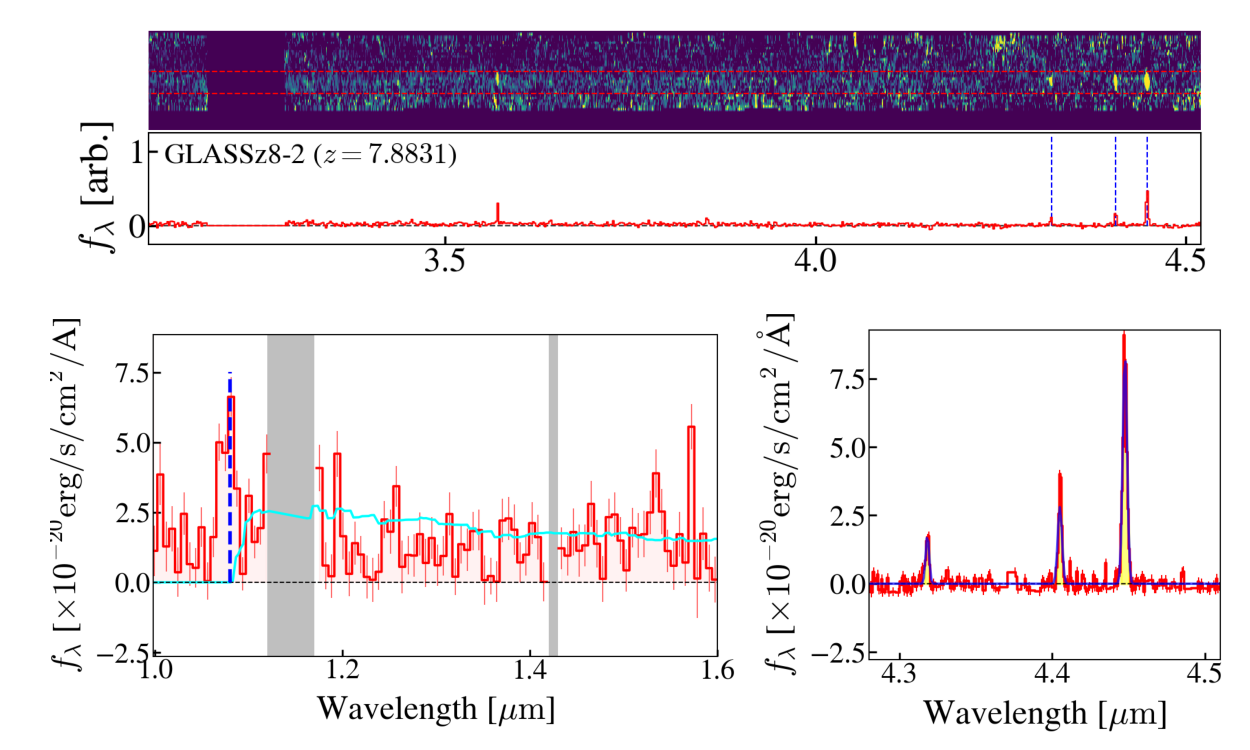}
	\caption{
	NIRSpec spectra in the observed wavelength frame (PRISM for the top four and G395H grating for the bottom three) of the confirmed protocluster members. For each galaxy, the top panel shows the 2D and 1D full spectrum, {where the position of Lyman break, \hb\ line, and \oiii\ doublets are indicated (blue vertical lines)}; the bottom left panel shows a zoomed region around the Lyman break, {along with a best-fit template by \gsf (cyan)}, where the detector gap region in the G140H grating is masked (gray); the bottom right panel shows the continuum-subtracted spectrum in the region of the \hb+\oiii-doublet lines, where lines with $>5\,\sigma$ detection are hatched in yellow. For YD4, where both PRISM and G395H spectra available, the latter is shown. The fitted three-component Gaussian model is also shown (blue). It is noted that the 1D full spectrum in the top panel is resampled to a coarse spectral grid for visualization purposes. 
    }
\label{fig:spec}
\end{figure*}

%%%%%%%%%%%%%%%%%%%%%
\section{Analysis and Results}\label{sec:ana}

%%%%%%%%%%%%%%%%%%%%%
\subsection{Spectroscopic analysis of $z\sim8$ candidates}\label{sec:spec}
We present our spectroscopic analyses of the \nspec\ galaxies in our sample in Fig.~\ref{fig:spec}, which shows the two-dimensional spectra and one dimensional extraction. Remarkably, all galaxies show clear \oiii\,5007 lines at $\sim4.4\,\mu$m, and tentative \hb, \oii, and [Ne~III] lines in a few galaxies (see \citealt{mascia23} and Roberts-Borsani et al., in prep. for a dedicated analysis of the emission lines). Here we focus on redshift determinations using the \oiii~4959,5007-doublet and \hb\ line.

The redshift of each source is determined by fitting a three-component Gaussian to \hb\ and the \oiii-doublet after subtracting the underlying continuum. Before subtraction, we first scale the observed spectrum by matching the continuum level to the best-fit model, for slitloss and any remaining  offset in absolute flux calibration. We use the wavelength range at 4-4.8\,$\mu$m while emission lines (i.e. \hb\ and \oiii) are masked. For continuum subtraction, we use a best-fit spectral energy distribution (SED) template derived with broadband photometry (Sec~\ref{sec:sed}). We fix the line ratio of the \oiii-doublet lines to 1:3 and set the width to a single parameter for the two components of the doublet. For \hb, the amplitude and line width are set as free parameters. Including the redshift, we have five free parameters. The uncertainties and parameter posterior distribution functions are estimated by sampling the parameter space via \texttt{emcee} \citep{foreman13}.

In order to assess the detection significance of \hb\ and \oiii\ emission, we estimate the noise level in the spectrum from 3.6--4.8\,$\mu$m. We measure total fluxes at various wavelengths integrated over $\pm2\,\sigma$ (where $\sigma$ is the best-fit Gaussian width of each emission line). We then compare the standard deviation of these fluxes (i.e., noise) with the emission lines integrated over $\pm2\,\sigma$ around the central wavelength. This test indicates secure ($>5\,\sigma$ confidence) detections of \oiii$_{\lambda5007}$ in all \nspec\ photometric candidates that were targeted, along with \oiii$_{\lambda4959}$ in five and \hb\ in three. The resulting line fluxes and spectroscopic redshifts are presented in Table~\ref{tab:source}. Total line fluxes are measured by integrating the best-fit gaussian component for each line when detected at $>5\,\sigma$. {It is noted that the slit loss is corrected by multiplying a median ratio of the best-fit SED (Sec.~\ref{sec:sed}) and the observed spectrum at 4.0-4.8\,$\mu$m after masking the wavelengths of} \hb\ and \oiii-doublet lines. The line fluxes and spectroscopic redshifts are presented in  Table~\ref{tab:source}.

{For the four galaxies observed with the prism configuration (YD4, YD7, YD8, and ZD6) and one with the high-resolution grating (GLASSz8-2), we have spectroscopic coverage at the wavelength of \ly, allowing us independent check of the inferred redshift measurements. Indeed, we confidently detect the \ly\ break at the expected wavelength for the redshift derived above for all of the four galaxies (Fig.~\ref{fig:spec}). For the other two galaxies (ZD2 and ZD3), while a small part of the wavelength range near \ly\ falls in the detector gap, the break is still consistent with the inferred redshift.}

%%%%%%%%%%%%%%%%%%%%%
\subsection{Spectral energy distribution}\label{sec:sed}
We analyze the SED of the individual galaxies by using photometric data that covers 0.4--5\,$\mu$m. We use the SED fitting code {\tt gsf} \citep{morishita19}, which allows flexible determinations of star formation histories in a non-parametric form, by finding an optimal combination of stellar and interstellar medium (ISM) templates. We generate templates of different ages, [1, 3, 10, 30, 100, 300]\,Myrs, {and metallicities $\log Z_*/Z_\odot \in [-2:0.4]$ at an increment of 0.1}, by using {\tt fsps} \citep{conroy09fsps,foreman14}. {A nebular component (emission lines and continuum)} that is characterized by an ionization parameter {$\log U \in [-3 : -1]$} is also generated by {\tt fsps} {\citep[see also][]{byler17}} and added to the template after multiplication by an amplitude parameter. Dust attenuation and metallicity of the stellar templates are treated as free parameters during the fit, {whereas the metallicity of the nebular component is fixed to the same value of the stellar component during the fitting process.}

The posterior distribution function of the parameters is sampled by using \texttt{emcee} for $10^5$ iterations with the number of walkers set to 100. The final posterior is collected after excluding the first half of the realizations (known as burn-in). The resulting physical parameters are quoted as the median of the posterior distribution, along with the 16\,th to 84\,th percentile uncertainty ranges. The star-formation rate is calculated by averaging the last 100\,Myr of the posterior star formation history. The inferred physical properties are presented in Table~\ref{tab:prop}.

To supplement our characterization of the overdensity (Sec.~\ref{sec:sd}), we include the remaining eight photometric candidates presented by \citet[][Fig.~\ref{fig:mosaic}]{zheng14}. As revealed by the spectroscopy, the confirmed sample consists both of ZDs and YDs. The ambiguity is likely due to the fact that the redshift of interest falls in the middle of the effective redshift ranges probed by the two color selections, which define $7<z<8$ and $8<z<9$ samples, respectively. 

We fit the redshifts of the remaining photometric candidates with {\tt EAzY} \citep{brammer08}. To exclude possible outliers, we only include those where the $2\,\sigma$ redshift uncertainty overlaps with $z=7.88$. After this selection, we have nine photometric sources. The SEDs of the selected photometric sources are fitted as described above, with the redshift fixed to $z=7.88$ (Table~\ref{tab:prop}).

\begin{figure*}
\centering
	\includegraphics[width=0.48\textwidth]{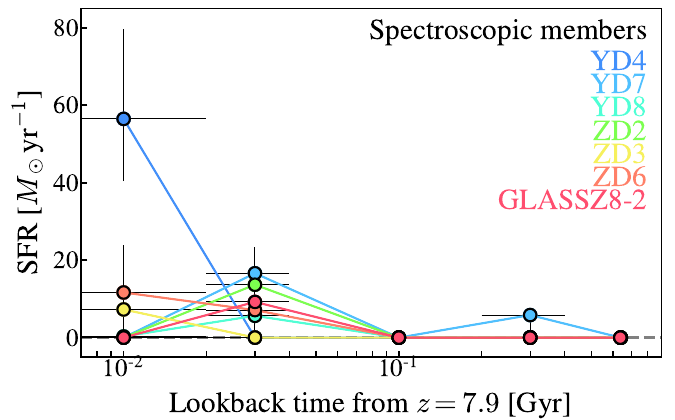}
	\includegraphics[width=0.48\textwidth]{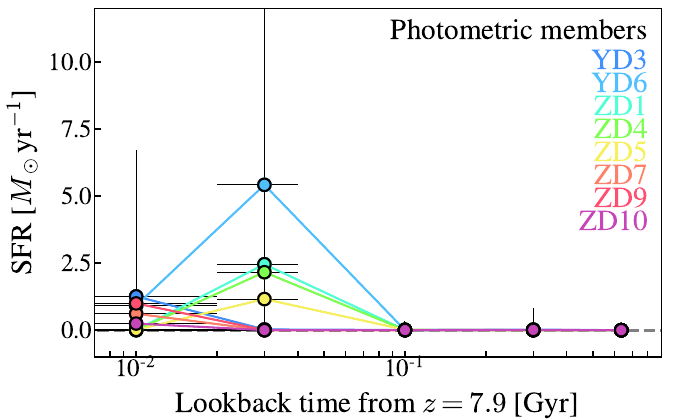}
	\caption{
    {Posterior star formation histories of the confirmed protocluster members (left panel) and phot-$z$ sample (right), estimated by {\tt gsf}. Most of those galaxies experience significant star formation in the last $<100$\,Myr, except for YD7, which is the most massive object among them. Star formation rates are corrected for magnification.}
    }
\label{fig:sfh}
\end{figure*}

{In Fig.~\ref{fig:sfh}, we show the derived star formation histories of the confirmed member galaxies and the phot-$z$ sample. All galaxies experience the peak of star formation in the last $<100$\,Myr. The exception is YD7, which formed about a half of the total mass at $\sim300$\,Myr prior to the observed redshift, making it a relatively old (with mass-weighted age $t_*\sim200$\,Myr) and massive system.}

{
Overdense regions are generally known as a place of accelerated evolution \citep[e.g.,][]{dressler80,thomas10}. It is thus of particular interest to investigate if there exists a luminous galaxy or quasar ($M_{\rm UV}<-22$) in/near an overdense region at high redshift. The sources presented here are relatively faint in rest-frame UV, with UV absolute magnitude, $M_{\rm UV}$, ranging from $-15.3$ to $-20.1$\,mag. On the other hand, some of the sources have moderate dust attenuation ($\simgt0.5$\,mag). Among those, YD4 is the most significant case with $A_V=1.1$, which is consistent with the $\sim4\sigma$ detection in dust continuum at its position \citep[][see also Sec.~\ref{ssec:previous}]{laporte17}. We compare $A_{V}$ values for our sample to those of a reference field sample that consists of 13 photometric sources at $z\sim8$ from \citet{nicha22}. While the distribution of our sample is skewed toward higher $A_V$, the shift is not statistically significant. We find no significant differences in other properties.
}

{We note that spectroscopic data are not included in the SED fitting process above; despite, the predicted line fluxes of the best-fit SED model overall show good consistency with the values measured in Sec.~\ref{sec:spec}. The only exception is YD8, where the observed flux is $\sim5\times$ larger than what is predicted by the best-fit model for both \hb\ and \oiii. While detail investigation on this discrepancy is deferred to future work, we note that the emission line templates used here are optimized for standard stellar populations and do not include extreme components such as, e.g., AGN.}

%%%%%%%%%%%%%%%%%%%%%
%%%%%%%%%%%%%%%%%

%%%%%%%%%%%%%%%%%%%%%
\subsection{Direct Estimate of Ionizing Photon Efficiency}\label{sec:xi}
{Having direct measurements of optical emission lines brings us not only solid redshift confirmation but also direct insight into ionizing properties of the ISM. One of such measurements is the ionizing photon production per unit UV luminosity, or ionizing photon efficiency, 
\begin{equation}
\xi_{\rm ion} = {\rm N_{\rm LyC}} / L_{\rm UV,1500}\,{\rm [Hz\,erg^{-1}]}
\end{equation}
where ${\rm N_{\rm Lyc}}$ is the total ionizing photons of Lyman continuum and $ L_{\rm UV,1500}$ intrinsic UV luminosity density measured at rest-frame 1500\,\AA\ \citep[e.g.,][]{schaerer16,Prieto-Lyon22}. 
The production efficiency is an important parameter, as the total ionizing output of galaxies can be simply parameterized \citep[e.g.,][]{madau99,robertson13} as the product of $\xi_{\rm ion}$ and the fraction of ionizing photons which escape the interstellar medium into the intergalactic medium, $f_{\rm esc}$. The direct measurement of $N_{\rm LyC}$ is not available for the redshift range of interest since the luminosity in the optical hydrogen recombination lines is proportional to the number of LyC photons absorbed in the galaxy. As a proxy, we adopt the following formula:
\begin{equation}
N_{\rm LyC} = 2.1 \times 10^{12} (1 - f_{\rm esc})^{-1} L({\rm H}\beta)_{\rm unatten.}
\end{equation}
The intrinsic UV luminosity density is inferred from the best-fit SED. We correct the measured \hb\ flux for attenuation by using the dust attenuation from the same SED modeling and assuming $E(B-V)_{\rm neb}/E(B-V)_{\rm stel.}=2.27$ \citep{shivaei20}. Using the two equations above, we derive the production rate of ionizing photons which did not escape the galaxy, $\xi_{\rm ion}(1-f_{\rm esc})$. The median value for the five objects available for the measurement is $\langle\xi_{\rm ion}(1-f_{\rm esc})\rangle=4.6\times10^{25}$\,erg Hz$^{-1}$. The measurements for individual galaxies are reported in Table~\ref{tab:prop}.
}

%%%%%%%%%%%%%%%%%%%%%
\section{Discussion}\label{sec:disc}

%%%%%%%%%%%%%%%%%%%%%
\subsection{Characterizing \targ: Estimate of size, overdensity, mass, and velocity dispersion}\label{sec:sd}

\begin{figure}
\centering
	\includegraphics[width=0.48\textwidth]{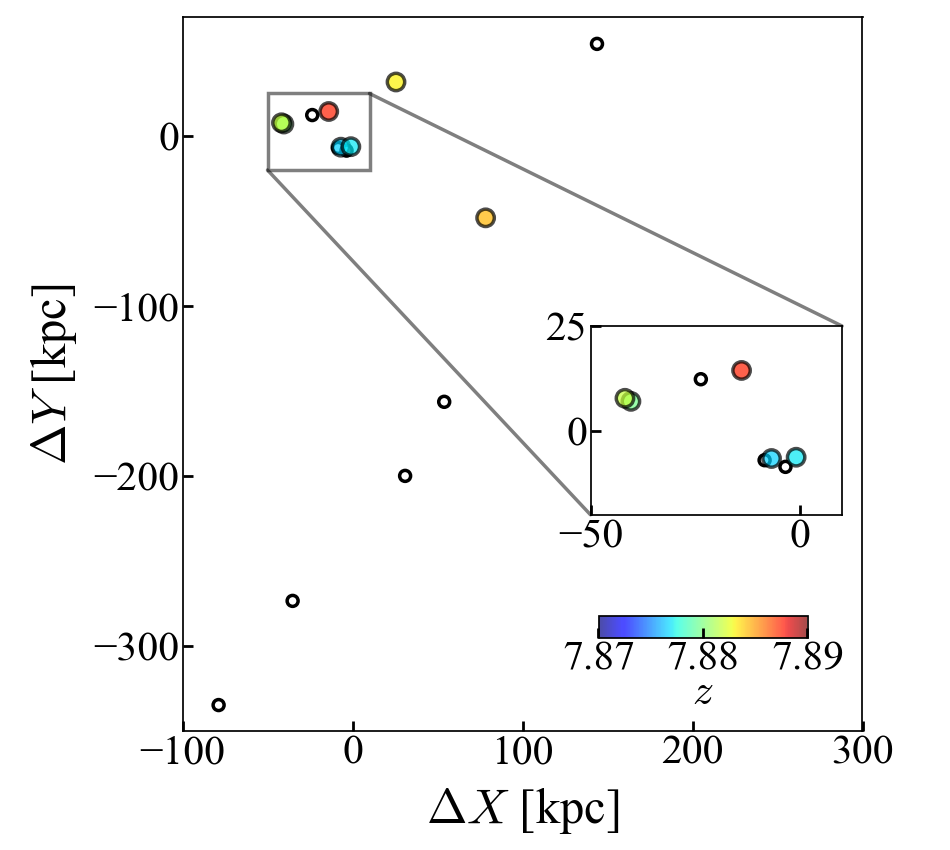}
	\caption{
	The source plane projected distribution of the confirmed protocluster member galaxies (circles, color-coded by spectroscopic redshift) and photometric candidates (open circles) in the proper scale. Positions of the galaxies are reconstructed on the source plane. The zoomed-in region around the overdensity is shown in the inset. The zeropoint of the coordinates is set to the centroid of the spectroscopic sources.
    }
\label{fig:3d}
\end{figure}

We first investigate the spatial distribution of the member galaxies in \targ. We use an updated version of the lens model presented by \cite{bergamini22} - which includes the recent spectroscopic confirmation of a triply-imaged $z\sim10$ LBG \citep{roberts-borsani22b} in the field (Bergamini et al. in prep.) - to correct for the magnification by the foreground cluster. The two-dimensional distribution of our sources in physical units is shown in Fig.~\ref{fig:3d}. After correcting for the lens magnification, we find that the confirmed sources are located within a circle of radius $R\sim 60$\,kpc in the source plane. The distribution on the sky is fairly elongated. 

{Based on the derived spatial distribution in the source plane, we estimate overdensity by including only spectroscopically confirmed members. For the field reference, we use the luminosity function at $z\sim8$ derived in \citet{bouwens21} and integrated it down to $M_{\rm UV}=-19$. Within a projected area of $r=60$\,kpc ($\sim12''$), we find $\bar{n}=0.3_{-0.1}^{+0.1}$, where the associated uncertainties reflect the 16--84\,th percentile ranges of the luminosity function adopted. This gives an estimate of the overdensity $\delta=24_{-8}^{+12}$. While our estimate is lower than the one derived by \citet[][$\delta=132_{-51}^{+66}$]{ishigaki16}, they included eight galaxies (including photometric candidates that are not confirmed in our study) in the central smaller region ($r=6''$). Therefore, given that only the spectroscopic sample is included, our estimate is likely a conservative lower limit.}

Secondly, we attempt to estimate the mass of the structure. 
Following previous work \cite[e.g.][]{laporte22}, we can estimate the halo mass of the individual components from the halo-mass galaxy-luminosity relation. Using the relation derived by \citet{mason22}, we infer that the brightest member of the overdensity (ZD2, $M_{\rm UV}=-20.1$) lives in a $M_{\rm h} \approx (7\pm 2) \times 10^{10}\,M_\odot$ halo. Summing the halo mass of all the confirmed members we obtain a lower limit to the total halo mass $\gtrsim 4\times10^{11}\,M_\odot$. 

Lastly, we can take advantage of the spectroscopic data to obtain for the first time an estimate of the velocity dispersion of a protocluster at such high redshift. Given the small number of measured redshifts we adopt a simple Gaussian estimator and bootstrap method to derive the uncertainty \citep{Beers90}, obtaining $1100\pm200\,\kms$. 
We caution the reader that the estimate should be treated with a degree of caution since the system is likely not virialized, and that in computing this quantity we are assuming the spread in redshift with respect to the mean is due to motion as opposed to distance along the line of sight. Nevertheless, we report it to assist future theoretical investigations.

%%%%%%%%%%%%%%%%%%%%%
%%%%%%%%%%%%%%%%%

%%%%%%%%%%%%
\subsection{Estimating the Present-day Mass of \targ}\label{sec:fate}
{With \nspec\ members being spectroscopically confirmed, it is of extreme interest to estimate the present-day halo mass of a system like \targ. We attempt to estimate the total cluster mass at $z=0$ by following the widely used formula
\begin{equation}\label{eq:Mz0}
    M_{z=0} = {\bar \rho} V_{\rm cor} (1+\delta_m)    
\end{equation}
 \citep[e.g.,][]{steidel98,chiang13}, where $V_{\rm cor} = \frac{V_{\rm obs}}{C}$ is the redshift-space distortion corrected comoving volume of the system, ${\bar \rho}$ the mean matter density of the Universe, and $\delta_m$ the mass overdensity. The correction coefficient $C$ and $\delta_m$ are linked as
\begin{equation}\label{eq:dm}
    1 + b \delta_m = C (1+\delta),
\end{equation}
where $b$ is the bias parameter and $\delta$ galaxy overdensity (Sec.~\ref{sec:sd}). By adopting the linear interpolation presented in \citet[][]{ouchi18}, we adopt the bias parameter $b = 6.5$ at $z=7.9$ of the total halo mass $1\times10^{11}\,M_\odot$. The correction coefficient is expressed as
\begin{equation}\label{eq:C}
    C = 1 + f - f(1+\delta_m)^{1/3},
\end{equation}
where $f$ is a function of the mass density parameter, $\Omega_M(z)$, at $z$
\begin{equation}
f = \Omega_M^{4/7}(z).
\end{equation}
By solving Eqs.~\ref{eq:dm} and \ref{eq:C}, we obtain $C=0.56_{-0.08}^{+0.08}$ and $\delta_m=2.00_{-0.48}^{+0.55}$ for the galaxy overdensity measurement estimated in Sec~\ref{sec:sd}. For the same area used in Sec~\ref{sec:sd} and the redshift interval $\delta z=0.011$, we estimate the comoving volume of \targ\ to be $V_{\rm obs}=9920$\,cMpc$^3$. By substituting this in Eq.~\ref{eq:Mz0}, we obtain $M_{z=0}=2.2_{-0.6}^{+0.9}\times10^{15}\,M_\odot$. This implies that \targ\ would be expected on average to become a Coma-like system at $z=0$, whereas $10^{14}\,\mathrm{M}_\odot$ is typically used as the threshold for a system to be called a ``cluster'' \citep[e.g.,][]{rosati02,overzier16}.

Another way of predicting the present-day mass is to find and compare with overdense systems like \targ\ in a simulation. Here we look into the EAGLE 100 Mpc ``Reference'' simulation \citep{schaye15}. We first extract all galaxies in the simulation with $M_* > 10^8\,M_\odot$ at $z = 8$ ($N = 557$), and trace them to their $z = 0$ descendants. To limit our analysis to the likely analogs of \targ, we trace overdense regions that have six or more galaxies in a spherical region of $r=300$\,pkpc at $z=8$ ($N=19$). At $z = 0$, the descendants of these 19 galaxies are hosted by halos in the mass range $\log M_{200c}/\mathrm{M}_\odot=13.5$-$14.5$, which is consistent with a model overdensity identified by \citet{ishigaki16} in their cosmological simulation. While this indicates that \targ\ could evolve into a system at the lower bound of typical clusters or even a group, it should be noted that the upper side of the resulting mass distribution above is likely limited by the simulation volume, as there is no object with $\log M_{200c}/M_\odot>14.5$ in the entire simulation box. Estimating a precise mass of the structure and its future evolution would require simulations with sufficient resolution and astrophysical detail to resolve individual galaxy components matched in luminosity or stellar mass, while simultaneously probing sufficient volumes, to include multiple structures of this kind to average out the expected stochasticity \citep{chiang13}, or mapping to a typical dark matter halo rarity then followed across cosmic time \citep{trentisantos2008}. A dedicated study is beyond the scope of this paper and left for future work.

Our very first attempt of spectroscopic followup on the bright members in the core region already confirmed \nspec\ member galaxies {\it at $100\,\%$ success rate}. For further characterization of the confirmed overdensity, a sample of galaxies at larger extent would be required. The progenitors of massive clusters are typically spread over several Mpc and thus to robustly estimate the mass of the descendant one would require a survey covering a much larger area of the sky \citep[e.g.,][]{overzier09,contini16}. In fact, we found in the simulation that the $z=0$ mass distribution skews at a higher mass when the search region is defined by a larger radius ($\sim2$\,pMpc) that contains more numerous galaxies ($N=35$ or more).
}

%%%%%%%%%%%%%%%%%
\subsection{On the absence of Lyman~$\alpha$ emission lines}\label{sec:bubble}
The absence of strong \ly\ emission provides insight into the intergalactic medium (IGM) properties surrounding the protocluster. None of our spectroscopically confirmed sources shows a clear \ly\ line (Fig.~\ref{fig:spec}). To quantify the non-detections, we estimate the upper limit on rest-frame equivalent widths of the line, EW$_0$(\ly), following \citet{hoag19} and \citet{morishita20}:
\begin{equation}
{\rm EW_0(Ly\alpha)} = \frac{F_{\rm Ly\alpha}}{f_{\rm cont.}} \frac{1}{(1+z)}
\end{equation}
For $f_{\rm cont.}$ we use the continuum model derived from our SED fitting analysis.  We replace the non-detected \ly\ flux with the limiting flux estimated over the instrumental resolution ($\sim100$\,\AA, or $2700\,\kms$) at the wavelength of \ly. The resulting range of upper limits is $\sim16$--28\,\AA\ ($2\,\sigma$; Table~\ref{tab:source}). The two galaxies (ZD2, ZD3) observed in the ERS programs are not available for the EW measument as the wavelength range of interest falls in the detector gap.

The lack of strong \ly\ emission is perhaps not surprising given the redshift $z=7.9$ of the host galaxies, where inferences on the ionization state of the IGM find neutral fractions in excess of $x_{\rm HI} > 70\,\%$ \citep{Mason2019,Hoag2019}. 
From the measured \ly\ EW limits of the four galaxies, and their absolute magnitudes, we estimate {the volume-averaged neutral hydrogen fraction of IGM}
%along the line of sight 
to be $>0.45$ (at 68\% CL) using the same methodology presented by \citet{Mason2018}. This is consistent with previous work on the cosmic average, within the uncertainties \citep{Mason2019}. We note that this analysis assumes the observations are independent sightlines. A more realistic analysis including their correlation within the same physical region would likely recover a slightly lower limit, but is beyond the scope of this work. A larger number of spectroscopic measurements in the protocluster are needed to refine this limit and identify potential differences with regard to the cosmic average.

In such a highly neutral environment large ionized bubbles are expected to be extremely rare \citep[e.g.,][]{Mesinger2007}. Even around regions of comparable overdensity containing sources of similar magnitude ranges, large reionization simulations predict median bubble sizes to be smaller than 1\,pMpc where $x_{\rm HI}>70\,\%$ (Lu et al. in prep, using the Evolution of Structure reionization simulations, \citealt{Mesinger2016}). 
If the bubble size is below $\sim1$\,pMpc the redshift along the line of sight is not sufficient for \ly~ to escape, and in fact \ly\ transmission is $\simlt20\,\%$ at its line center \citep{mason20,qin21}. 
{Assuming a gaussian-shape emission line with velocity offset of \ly\ from systemic of $200$\,km\,s$^{-1}$ (which is likely an overestimate in these low luminosity galaxies, \citealt{Mason2018}) and FWHM equal to the velocity offset, the total fraction of transmitted \ly\ flux is expected to be $< 40\%$ for a 1\,pMpc bubble and $< 30\%$ for a 0.75\,Mpc bubble. Assuming the average \ly\ EW\,$=30 {\mathrm \AA}$ for $M_{\rm UV}\sim-19.75$ galaxies at $z\sim6$ from \citet{deBarros17} as the ``emitted" EW, we would thus expect to observe \ly, after transmission through the IGM, with EW\,$ < 12 {\mathrm \AA}$, below our detection threshold (Table~\ref{tab:source}).
}

We can verify our theoretical expectation by
estimating the radius of an \hii\ region, $R_{\rm HII}$, ionized by UV photons of a single galaxy using the equation in \citet[][also \citealt{endsley22}]{haiman97}:
\begin{eqnarray} \label{eq:RHII}
    \frac{dR_{{\mathrm{HII}}}}{dt} &=& \frac{\langle\xi_{\rm ion}{}\rangle\ \langle f_{\rm esc}{}\rangle\ L_{\mathrm{UV}}}{4 \pi R_{_{\mathrm{HII}}}^2\ \bar{n}_{{\mathrm{HI}}}(z)} + R_{_{\mathrm{HII}}}\ H(z) \nonumber \\
    && - R_{_{\mathrm{HII}}}\ \alpha_B\ \bar{n}_{{\mathrm{HI}}}(z)\ \frac{C_\mathrm{HI}}{3}.
\end{eqnarray}
For simplicity, this equation assumes that the ionizing bubble is spherically symmetric and created by a single source at its center. We adopt a Case~B recombination coefficient, $\alpha_{B}=2.59\times10^{-13}$\,cm$^3$/s \citep{osterbrock89}, and ionizing photon escape fraction of $\langle f_{\rm esc}\rangle = 0.2$ and an IGM \hi\ clumping factor $C_{\rm HI}=3.0$ \citep{shull2012,robertson13}. {For the ionizing photon production efficiency, we use those derived in Sec.~\ref{sec:xi} for those with \hb\ flux measurements available. We adopt the median value for the remaining sample.}

By using the derived star formation history and luminosity presented in Sec.~\ref{sec:sed}, we estimate the bubble size for the spectroscopically confirmed individual sources and for the photometrically-selected sample identified in Sec.~\ref{sec:sed}. The estimated bubble sizes are $\ll1$\,Mpc for most of the sample, due to the relatively low UV luminosity of the galaxies (Table~\ref{tab:prop}). Given that their separation in the source plane is of order 60~pkpc, we also estimate the bubble size by considering the cumulative effect of all the confirmed sources as if they were colocated. Even in this case, we find it to be $R\sim0.78$\,Mpc, i.e. insufficient to allow \ly\ to escape. Even the inclusion of all photometric candidates is not significantly changing the estimate, as those are mostly fainter than the confirmed members (also see their individual estimates in Table~\ref{tab:prop}).

In conclusion, our bubble size estimate is consistent with the non-detection of  \ly\  in our confirmed sources. By comparison, at slightly lower redshifts, \citet{endsley22} estimated bubble sizes of $0.7$--1.1\,Mpc for UV-bright ($-M_{\rm UV}\sim20$--$22$\,mag) galaxies at $z=6.6$--6.9. %Among 10 of their primary sample, 
\ly\ was detected in nine out of ten galaxies, showing that both an overdense environment and sufficient ionizing photon flux is required to produce an ionized bubble large enough to allow significant transmission of \ly\ photons (c.f. the comparison of Ly$\alpha$ detections in UV-bright and fainter galaxies in \citealt{rb22}).

\subsection{Comparison with previous work}
\label{ssec:previous}
Our estimated lower limit of the mass of \targ\ is comparable to those of previously known protoclusters and protoscluster candidates using similar methods, including the recently reported candidate behind the SMACS0723 cluster \citep{laporte22}, {where two of the member candidates are spectroscopically confirmed to be $z=7.66$.} The real breakthrough of our observations, however, is the sheer number of spectroscopically confirmed redshift measurements, which allow us to establish secure membership to the protocluste and get a first estimate of the its velocity dispersion. This clearly provides a glimpse of the power of JWST to add unprecedented detail to studies of the progenitors of today's large-scale structures.

\begin{figure}
\centering
	\includegraphics[width=0.48\textwidth]{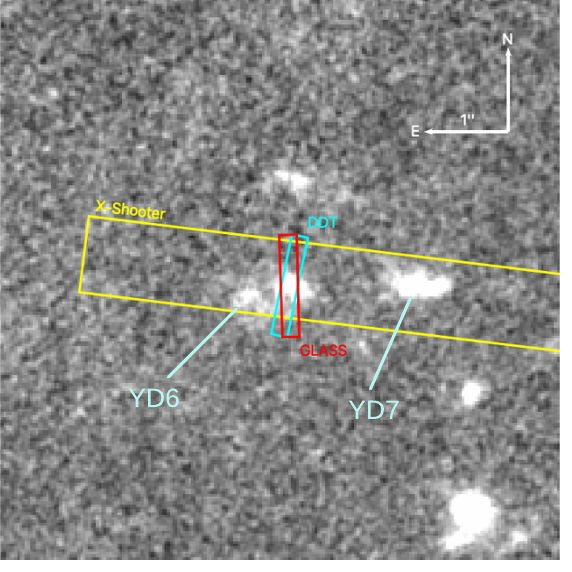}
	\caption{
	{A $5.4''\times5.4''$ smoothed F200W NIRCam image centered on YD4, with slits from various programs overlayed. The $\sim0.2''\times1.2''$ NIRSpec/MSA slits from GLASS and the DDT are shown in red and cyan, respectively, while the $\sim0.9''\times11''$ VLT/X-Shooter long-slit from \citep{laporte17} is shown in yellow. The two NIRSpec slits clearly isolate YD4 from a nearby, fainter companion $\sim0.5''$ South-East of its position (YD6), however the X-Shooter long-slit includes both objects, as well as YD7 located to its West.}
    }
\label{fig:slits}
\end{figure}

The spectroscopic redshift of \targ\ is in agreement with previous photometric redshift estimates, but in apparent tension with the redshift reported for YD4 ($z=8.38$) based on \ly\ and ALMA \cii\ 158\,$\mu$m and \oiii\ 88\,$\mu$m emission \citep{laporte17,laporte19,carniani20}. A likely explanation of the apparent tension is a line-of-sight superimposition of sources at similar redshifts. This is a common occurrence in the photometric identification of overdensities and should be kept in mind when considering protocluster candidates without spectroscopy. As shown in Figure~\ref{fig:slits}, YD4 is close on the sky (separation $\sim0.5''$) to a secondary $Y$-dropout source (YD6; \citealt{zheng14}), which falls outside our NIRSpec spectroscopic apertures (in the GLASS and DDT observations) but lies within the VLT/X-Shooter long-slit and indistinguishable from YD4 at ALMA resolution (in the case of \cii). We hypothesize that the source detected at $z=8.38$ is actually in the background of the protocluster and likely associated with YD6 \citep[estimated to be at $z_{\rm phot}=8.3\pm0.2$][]{zheng14}. In fact, the \oiii\ 88 $\mu$m flux appears better centered on the fainter counterpart while the detection of dust appears associated with YD4 (see Figure 2 in \citealt{laporte19}), consistent with both the large dust quantity {($A_V=1.1$) and red UV slope ($\beta_{\rm UV}=-1.3$) estimated for YD4 in Section \ref{sec:sed} and the discrepant spectroscopic redshifts.} More extensive spectroscopic coverage is needed to confirm the hypothesis.

%%%%%%%%%%%%%%%%%%%%%
%%%%%%%%%%%%%%%%%

%=============================
\begin{deluxetable*}{ccccccccccc}%[!h]
\tabletypesize{\footnotesize}
\tablecolumns{11}
\tablewidth{0pt} 
\tablecaption{Physical properties of the final candidates.}
\tablehead{
\colhead{ID} & \colhead{$\mu^{\dagger}$} & \colhead{$M_{\rm UV}$} &\colhead{$\beta_{\lambda}$} & \colhead{$\log M_*$} & \colhead{SFR} & \colhead{$\log t_*$} & \colhead{$\log Z_*$} & \colhead{$A_V$} & \colhead{$R_{\rm HII}^\ddagger$} & \colhead{$\xi_{\rm ion }(1-f_{\rm esc})$$^{\mathsection}$}\\
\colhead{} & \colhead{} &\colhead{mag} & \colhead{} & \colhead{$M_\odot$} & \colhead{$M_\odot$\,/\,yr} & \colhead{Gyr} & \colhead{$Z_\odot$} & \colhead{mag} & \colhead{pMpc} & \colhead{$10^{25}$\,Hz/erg}
}
\startdata
\cutinhead{Spectroscopic sample}
YD4 & $2.01_{-0.04}^{+0.05}$ & $-19.69_{-0.06}^{+0.08}$ & $-1.28_{-0.11}^{+0.07}$ & $9.17_{-0.17}^{+0.27}$ & $6.46_{-1.99}^{+1.94}$ & $-1.44_{-0.53}^{+0.69}$ & $-1.30_{-0.38}^{+0.42}$ & $1.10_{-0.15}^{+0.16}$ & $0.10_{-0.09}^{+0.31}$ & $4.06_{-1.36}^{+1.49}$\\
YD7 & $2.02_{-0.04}^{+0.05}$ & $-19.96_{-0.05}^{+0.05}$ & $-1.86_{-0.04}^{+0.02}$ & $9.39_{-0.15}^{+0.08}$ & $4.04_{-1.57}^{+1.51}$ & $-0.71_{-0.26}^{+0.22}$ & $-1.16_{-0.49}^{+0.37}$ & $0.25_{-0.14}^{+0.16}$ & $0.76_{-0.17}^{+0.11}$ & --\\
ZD6 & $1.94_{-0.04}^{+0.05}$ & $-19.39_{-0.10}^{+0.09}$ & $-1.53_{-0.07}^{+0.08}$ & $8.86_{-0.19}^{+0.27}$ & $3.34_{-1.08}^{+2.98}$ & $-1.41_{-0.38}^{+0.40}$ & $-0.86_{-0.69}^{+0.76}$ & $0.71_{-0.26}^{+0.23}$ & $0.26_{-0.17}^{+0.18}$ & --\\
YD8 & $2.63_{-0.06}^{+0.07}$ & $-19.28_{-0.05}^{+0.04}$ & $-2.25_{-0.05}^{+0.02}$ & $8.36_{-0.26}^{+0.09}$ & $1.34_{-0.66}^{+0.30}$ & $-1.48_{-0.27}^{+0.18}$ & $-0.43_{-0.42}^{+0.13}$ & $0.14_{-0.04}^{+0.06}$ & $0.53_{-0.18}^{+0.07}$ & $19.59_{-2.33}^{+2.59}$\\
ZD2 & $1.98_{-0.04}^{+0.05}$ & $-20.13_{-0.02}^{+0.02}$ & $-2.24_{-0.01}^{+0.01}$ & $8.73_{-0.05}^{+0.03}$ & $3.13_{-0.37}^{+0.24}$ & $-1.49_{-0.31}^{+0.16}$ & $-0.29_{-0.05}^{+0.08}$ & $0.20_{-0.01}^{+0.01}$ & $0.58_{-0.02}^{+0.03}$ & $10.98_{-0.61}^{+0.63}$\\
ZD3 & $1.94_{-0.04}^{+0.05}$ & $-19.67_{-0.07}^{+0.06}$ & $-2.07_{-0.03}^{+0.05}$ & $8.33_{-0.15}^{+0.32}$ & $1.00_{-0.28}^{+0.59}$ & $-1.44_{-0.47}^{+0.61}$ & $-0.64_{-0.68}^{+0.59}$ & $0.22_{-0.14}^{+0.14}$ & $0.07_{-0.05}^{+0.23}$ & $1.62_{-0.87}^{+0.59}$\\
GLASSZ8-2 & $2.15_{-0.04}^{+0.06}$ & $-19.75_{-0.04}^{+0.03}$ & $-2.18_{-0.02}^{+0.02}$ & $8.56_{-0.04}^{+0.05}$ & $2.13_{-0.18}^{+0.25}$ & $-1.46_{-0.32}^{+0.17}$ & $-0.21_{-0.05}^{+0.04}$ & $0.24_{-0.02}^{+0.02}$ & $0.38_{-0.02}^{+0.03}$ & $4.59_{-0.54}^{+0.58}$\\
\cutinhead{Photometric sample$^*$}
YD3 & $2.79_{-0.05}^{+0.08}$ & $-17.51_{-0.19}^{+0.16}$ & $-2.04_{-0.11}^{+0.09}$ & $7.78_{-0.28}^{+0.19}$ & $0.27_{-0.13}^{+0.18}$ & $-1.33_{-0.47}^{+0.43}$ & $-1.38_{-0.42}^{+0.59}$ & $0.39_{-0.26}^{+0.32}$ & $0.09_{-0.07}^{+0.18}$ & --\\
YD6 & $2.00_{-0.04}^{+0.05}$ & $-18.23_{-0.12}^{+0.11}$ & $-1.61_{-0.08}^{+0.06}$ & $8.61_{-0.27}^{+0.19}$ & $1.58_{-0.74}^{+1.28}$ & $-1.32_{-0.32}^{+0.57}$ & $-1.29_{-0.44}^{+0.83}$ & $0.69_{-0.35}^{+0.24}$ & $0.23_{-0.13}^{+0.16}$ & --\\
ZD1 & $2.02_{-0.04}^{+0.05}$ & $-17.08_{-0.34}^{+0.25}$ & $-1.64_{-0.13}^{+0.29}$ & $8.19_{-0.28}^{+0.28}$ & $0.70_{-0.37}^{+0.78}$ & $-1.35_{-0.27}^{+0.40}$ & $-1.14_{-0.61}^{+0.86}$ & $0.55_{-0.33}^{+0.51}$ & $0.17_{-0.05}^{+0.12}$ & --\\
ZD4 & $1.96_{-0.04}^{+0.05}$ & $-17.59_{-0.19}^{+0.20}$ & $-1.89_{-0.16}^{+0.09}$ & $8.09_{-0.28}^{+0.31}$ & $0.59_{-0.30}^{+0.60}$ & $-1.30_{-0.27}^{+0.49}$ & $-1.16_{-0.51}^{+1.04}$ & $0.35_{-0.23}^{+0.36}$ & $0.20_{-0.07}^{+0.13}$ & --\\
ZD5 & $3.30_{-0.13}^{+0.13}$ & $-17.55_{-0.19}^{+0.15}$ & $-2.13_{-0.04}^{+0.08}$ & $7.84_{-0.37}^{+0.27}$ & $0.36_{-0.20}^{+0.31}$ & $-1.29_{-0.33}^{+0.35}$ & $-0.84_{-0.62}^{+0.54}$ & $0.16_{-0.10}^{+0.21}$ & $0.20_{-0.11}^{+0.12}$ & --\\
ZD7 & $10.85_{-0.38}^{+0.39}$ & $-17.44_{-0.12}^{+0.10}$ & $-2.39_{-0.04}^{+0.03}$ & $7.27_{-0.30}^{+0.26}$ & $0.09_{-0.05}^{+0.07}$ & $-1.49_{-0.45}^{+0.52}$ & $-1.59_{-0.31}^{+0.48}$ & $0.10_{-0.08}^{+0.20}$ & $0.07_{-0.05}^{+0.13}$ & --\\
ZD9 & $4.28_{-0.11}^{+0.14}$ & $-17.53_{-0.12}^{+0.11}$ & $-2.08_{-0.02}^{+0.03}$ & $7.53_{-0.29}^{+0.42}$ & $0.14_{-0.07}^{+0.13}$ & $-1.42_{-0.48}^{+0.66}$ & $-0.72_{-0.90}^{+0.85}$ & $0.20_{-0.12}^{+0.24}$ & $0.07_{-0.04}^{+0.19}$ & --\\
ZD10 & $6.29_{-0.34}^{+0.27}$ & $-15.33_{-0.52}^{+0.39}$ & $-1.54_{-0.29}^{+0.44}$ & $7.50_{-0.37}^{+0.28}$ & $0.09_{-0.06}^{+0.11}$ & $-1.07_{-0.59}^{+0.46}$ & $-0.86_{-0.92}^{+0.85}$ & $0.55_{-0.43}^{+0.61}$ & $0.09_{-0.07}^{+0.16}$ & --\\
\enddata
\tablecomments{
$^\dagger$Median magnification of the lens model by Bergamini (in prep.), calculated at the position of the source. 
Measurements are associated with $1\,\sigma$ uncertainties.
$^\ddagger$HII bubble size, calculated with Eq.~\ref{eq:RHII}.
$^\mathsection$Ionizing photon production efficiency (Sec.~\ref{sec:xi}). $^*$Sources selected by \citet{zheng14} that were not observed in our NIRSpec programs. Only those with photometric redshift consistent with $z=7.88$ at $2\,\sigma$ are included. Redshift of the photometric sample is fixed to $z=7.88$ in the SED analysis.
{$t_*$: Mass-weighted age calculated over the posterior star formation history.} Star-formation rate is calculated by averaging the last 100\,Myr of the posterior star formation history. All measurements are corrected for magnification. 
}\label{tab:prop}
\end{deluxetable*}
%=============================

%=============================
\section{Summary}\label{sec:sum}

In this work, we presented a JWST NIRSpec spectroscopic follow-up analysis of \nspec\
photometrically-selected members of a galaxy overdensity in the epoch of reionization at $z=7.9$, leading to robust redshift measurements for all photometric candidates by detecting \oiii$_{\lambda5007}$ and other rest-frame optical lines. The spectroscopic confirmation of the member galaxies in the core region allowed us to estimate overdensity, $\delta=24_{-12}^{+8}$, which characterizes \targ\ one of the most extreme overdensities in the early universe, with a lower limit on its halo mass of $\gtrsim 4\times10^{11}\,M_\odot$. We also obtained a first estimate of the velocity dispersion of the system ($\sigma =1100\pm 200$ km~s$^{-1}$), which will aid to compare the observations to similar structures identified in cosmological numerical simulations. {By using an empirical relation, we estimated the present-day halo mass of \targ\ to be $M_{z=0}=2.2_{-0.6}^{+0.9}\times10^{15}\,M_\odot$, comparable to the Coma cluster. Our analysis using a cosmological simulation suggests that spectroscopic confirmation of additional member galaxies at a further distance ($\sim2$\,pMpc) will further secure the present-day mass estimate.}

Our results clearly show the incredible potential of JWST to confirm $z>7$ redshifts thanks to the multiplexing capabilities afforded by the NIRSpec MSA. Remarkably, our study reports a $100\,\%$ success rate in identifying the redshifts of candidates at high $z$ independently of Ly$\alpha$, adding a further five confirmed $z>7$ sources to the literature. Crucially, we refined a previously reported spectroscopic redshift for YD4, suggesting line-of-sight superposition of two distinct sources. This work showcases JWST's potential to open a window for determining the properties of galaxies in the early universe. In particular -- upon the acquisition of a sufficient sample both in \targ\ and field control sample at similar redshift -- of particular interest for future progress will be the environmental dependence of physical properties of the sources, which we addressed in Sec.~\ref{sec:sed} for the present sample. In turn, this will help understand the role of galaxy clustering during cosmic reionization.

%=============================
\section*{Acknowledgements}
We thank Richard Ellis and Nicolas Laporte for useful conversations on the interpretation of the spectroscopic redshift of YD4. We thank the anonymous referee for their careful reading of the manuscript and constructive comments.
Support for program JWST-ERS-1324 was provided by NASA through a grant from the Space Telescope Science Institute, which is operated by the Association of Universities for Research in Astronomy, Inc., under NASA contract NAS 5-03127. Some of the data presented in this paper were obtained from the Mikulski Archive for Space Telescopes (MAST) at the Space Telescope Science Institute. The specific observations analyzed can be accessed via \dataset[https://doi.org/10.17909/bwwe-4a42]{https://doi.org/10.17909/bwwe-4a42}. 
KG acknowledges support from Australian Research Council Laureate Fellowship FL180100060. CM and TYL acknowledge support by the VILLUM FONDEN under grant 37459. The Cosmic Dawn Center (DAWN) is funded by the Danish National Research Foundation under grant DNRF140. This research is supported in part by the Australian Research Council Centre of Excellence for All Sky Astrophysics in 3 Dimensions (ASTRO 3D), through project number CE170100013.
We acknowledges support from the INAF Large Grant 2022 “Extragalactic Surveys with JWST”  (PI Pentericci). We acknowledge financial support from grants PRIN-MIUR 2017WSCC32 and 2020SKSTHZ.P
MB acknowledges support from the Slovenian national research agency ARRS through grant N1-0238. YMB gratefully acknowledges funding from the Dutch Science Organisation (NWO) under Veni grant number 639.041.751.
A.V.F. is grateful for financial support from the Christopher R. Redlich Fund
and many individual donors.
BM acknowledges support from Australian Government Research Training Program (RTP) Scholarships and the Jean E Laby Foundation.
XW is supported by CAS Project for Young Scientists in Basic Research, Grant No. YSBR-062.
RAW acknowledges support from NASA JWST Interdisciplinary Scientist grants
NAG5-12460, NNX14AN10G and 80NSSC18K0200 from GSFC.

%{{\it Software:} Astropy \citep{astropy:2013, astropy:2018}, numpy \citep{oliphant2006guide,van2011numpy}, python-fsps \citep{foreman14}, EMCEE \citep{foreman13}. }


\bibliography{sample631}{}
\bibliographystyle{aasjournal}



\end{document}